\documentclass{emulateapj}

\newcommand{\ha}{H$\alpha$}
\newcommand{\oiii}{[\ion{O}{3}]}
\newcommand{\ovi}{\ion{O}{6}}
\newcommand{\nev}{[\ion{Ne}{5}]} 

\shortauthors{Rakowski, Raymond, \& Szentgyorgi}
\shorttitle{[Ne V] Imaging of N49}
\submitted{submitted July 28, 2006, accepted September 15, 2006}
\journalinfo{ApJ accepted September 15, 2006}
\begin{document}

\title{ [Ne V] Imaging of N49 in the Large Magellanic Cloud}


\author{Cara E. Rakowski}
\email{crakowski@cfa.harvard.edu}

\author{John C. Raymond}


\author{Andrew H. Szentgyorgyi}
\affil{Harvard-Smithsonian Center for Astrophysics}

\begin{abstract}

We present sub-arcsecond imaging in \nev\ of N49, the brightest
optical SNR in the LMC. Between the ``cool'' optical and
``hot'' X-ray regimes, \nev\ emission indicates intermediate
temperatures for collisionally excited plasmas  (2--6$\times 10^{5}$K), 
for which imaging has been extremely limited.  
We compare the flux in these images to the \ovi\ measured spectroscopically
by FUSE in individual  apertures and find dereddened line ratios that
are reasonably consistent with our predictions for intermediate
velocity shocks. The overall luminosity in 
\nev\ for the entire remnant is $1.2\times 10^{36}~\rm erg~s^{-1}$,
which, given the measured line ratios, implies an overall \ovi\ luminosity of
$1.5 \times 10^{38}~\rm erg~s^{-1}$. 
These large radiative losses indicate that this material must have been
shocked recently relative to the total lifetime of the remnant.
We also explore the complex spatial structure. 
We find \nev\ tracing the \oiii\ emission more closely than
it does \ha,
measure significant shifts ($\sim$0.1 pc) 
between the peaks of different emission lines, and find two orders of
magnitude variations in the flux ratios for different filaments across
the remnant. These properties as well as the general filamentary
character of the optical emission suggest thermally
unstable intermediate velocity shocks.

\end{abstract}

\keywords{supernova remnants:individual(N49) --- ISM: structure --- shock waves --- ultraviolet: ISM}

\section{Introduction}

The ultraviolet (UV) transitions of \nev\ and \ion{O}{6} probe 
intermediate temperatures for collisionally
ionized plasmas [(2--6)$\times 10^{5}$ K] inaccessible 
in the optical or X-ray wavebands. This virtually unexplored
temperature range has the most efficient radiative cooling and dominates the
radiative losses in supernova remnants (SNRs).  Of the potential
diagnostics in this regime, the forbidden \nev\ lines $\lambda\lambda$
3345.8, 3425.9\AA\ are the only bright emission lines accessible from the
ground.  

While the Far-Ultraviolet Spectroscopic Explorer (FUSE) has
successfully provided emission line spectra from intermediate
temperature SNR-shocks
\citep{2001ApJ...549..416S,2002ApJS..140..367B,2002ApJ...565..297S,2004AJ....128.1615S,2004ApJ...615..280K}
it is not an imaging instrument. Study of the spatial
distribution of this intermediate temperature material has been extremely
limited. Low spatial resolution ($\sim$10\arcmin) \ovi\ images of the
Cygnus Loop and the Vela SNR have
been taken above the atmosphere with the High Resolution Emission Line
Spectrometer (HIRES) \citep{1992ApJ...396L.103R}
and SPEAR (Spectroscopy of Plasma Evolution from Astrophysical
Radiation) \citep{2006ApJ...644L.171N} respectively. 
However, the only high spatial resolution ($<1$\arcsec) image to date
of a SNR is the ground-based \nev\ observation of the Cygnus Loop
\citep{2000ApJ...529..279S}. 
Ground based Fabry-Perot observations of 64 individual sites within
the Cygnus Loop in \nev\ that incorporate spectral information from
the tuneable filter were conducted by \citet{1999A&A...351..669S} but
at much lower spatial resolution and hence lower sensitivity to sharp
features. 

\object[SNR 0525-66.1]{N49} is a good target for exploratory \nev\
imagery for a number of reasons. It is the brightest optical SNR
in the Large Magellanic Cloud (LMC), and hence has a known distance, is 
likely to be detectable, and is very well-studied. Previous work has
shown evidence of both radiative and hot X-ray emitting shocks. 
\citet{1985MNRAS.212..799S} interpreted this in terms of 
the evolution of a SNR blast-wave into a complex circumstellar
environment sculpted by the progenitor star's radiation and winds 
propagating through an existing multi-phase interstellar medium. 
Given the presence of a much larger existing molecular cloud
coincident with the south-east portion of N49
\citep{1991supe.conf..679H}, \citet{1992ApJ...394..158V} propose a
simpler picture in which the temperature and velocity structure seen
in N49 are the direct consequence of a Sedov-like blast-wave running into 
existing large clouds. The filamentary structure of the optical image
and the bands seen in the long-slit echelle data
would both be the natural consequence of a ``rippled sheet'' geometry
\citep{1987ApJ...314..187H} of alternating
face-on and edge-on shocks as the blast wave envelopes the large
clouds. 
\citet{1992ApJ...394..158V} derived a pre-shock density of
20--940~cm$^{-3}$ which they
found to be inconsistent with Shull's model and state that their 
geometry better explains the structure of the echelle data and
the fluxes seen.

\begin{deluxetable*}{lccccc}
\tablecaption{Observations
\label{observe} }
\tablewidth{0pt}
\tablehead{
Filter & Central $\lambda$ & FWHM & Exposure Time & Airmasses & 
Conversion\tablenotemark{a}\\
  &  (\AA)  &  (\AA) & (s) & & erg cm$^{-2}$ counts$^{-1}$
}
\startdata
\nev\   & 3426 & 15 & 1200 &  1.35,1.38,1.42 &  4.14E-15 \\
UV continuum & 3388 & 35 & 1200 &  1.48,1.54,1.60 & \\
\ha\           & 6563 & 46 & 100 &  1.33,1.68,1.69,1.70 & 6.18E-17 \\
red continuum & 7130 & 58 & 25 & 1.34,1.72,1.72,1.72 & \\
\oiii\  & 5007 & 51 & 50 & 1.73,1.74,1.74, & 1.14E-16 \\
  &  & & & 1.77,1.79,1.80,1.81 &  \\
green continuum & 5260 & 59 & 20 & 1.82,1.84,1.86 &
\enddata
\tablenotetext{a}{Conversion from counts in median combined, continuum
  subtracted final images}
\end{deluxetable*}

\section{Observations, Data Reduction and Analysis}

Observations of N49 were conducted on 2004 February 16, at the Clay
Telescope, one of the twin 6.5m Magellan telescopes at Las Campanas
Observatory, Chile (-29\arcdeg\ latitude) using the Raymond and Beverly
Sackler Magellan Instant Camera (MagIC). The CCD is a grade 0 SITe
2048$\times$2048 pixel array cooled to a detector temperature of
-120\degr\ C.  The 24\micron\ pixels each subtend 0.069\arcsec\ and the
seeing varied between 0.\arcsec6 and 0.\arcsec8. At 50 kpc to the LMC
each pixel corresponds to 0.0167~pc.
We installed custom \nev , \ha , \oiii , and associated continuum
filters manufactured by  Omega Optical Inc. for our
observations (see Table \ref{observe}). The \nev\ filter was narrower and the
UV-continuum wider than those used at Whipple Observatory for the study
of the Cygnus Loop \citep{2000ApJ...529..279S}.

SNR N49 was at an air mass less than 1.5 for the first half of the
night before 4:00UT, during which time we concentrated on the \nev\
observations. The \ha\ and \oiii\ images were taken primarily to
facilitate comparisons with \nev.
Table~\ref{observe} lists the relevant details of the observations.
Routine processing of the data was conducted in IRAF\footnote{IRAF is
  distributed by the National Optical Astronomy Observatories, which
  are operated by the Association of Universities for Research in
  Astronomy, Inc., under cooperative agreement with the National
  Science Foundation}  
making use of the MagIC package for the specifics of the CCDs. Flat-fielding
was performed using twilight sky flats and cosmic-rays were eliminated
by median-combining the separate frames. 
Given the extremely low
surface brightness in \nev\ and the UV continuum, high-frequency
electrical noise in the CCD read-out which produced a ``herringbone''
pattern of 5 to 10 counts could not be ignored. 
However, due to its slowly temporally varying periodicity it was possible to
greatly reduce or even eliminate this signal relative to the diffuse
target emission.\footnote{A series of ``PyRAF'' scripts to identify
  and remove the pattern of electrical noise that combine the IDL and
  IRAF procedures written by S. Burles and M. Holman, is available on
  request to C.~E.~Rakowski. However owing to differences in IRAF and
  Python distributions and settings, ``debugging'' is likely to be
  required.}

\begin{figure*}
\noindent \includegraphics[width=7.0in]{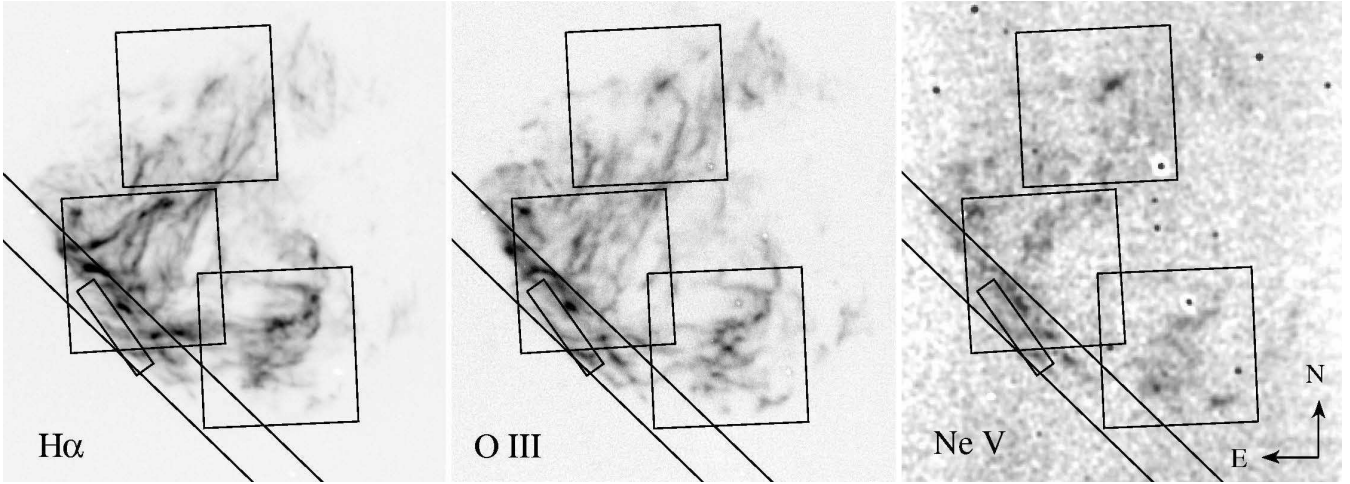}
\figcaption{Emission-line images of N49 \ha\ (\textsl{left}) \oiii\
  (\textsl{middle}) \nev\ (\textsl{right}). HUT and FUSE apertures are
  overlaid. The HUT aperture is the large diagonal swath across the
  south-east while the 4\arcsec $\times$ 20\arcsec\ diagonal aperture is
  FUSE C05501. The 30\arcsec\ $\times$ 30\arcsec\ apertures are
  C05502, X00502, and C05503 from north to south.
  The images are linearly scaled over somewhat arbitrary ranges chosen
  to capture the full dynamic range of each image. 
  The \nev\ image was first binned by 4 pixels then
  smoothed with a 0.828\arcsec\ Gaussian kernel. \ha\ and \oiii\ were
  not binned or smoothed.
\label{fuseap}}
\end{figure*}
\begin{figure*}
\begin{center}
\noindent \includegraphics[width=6.0in]{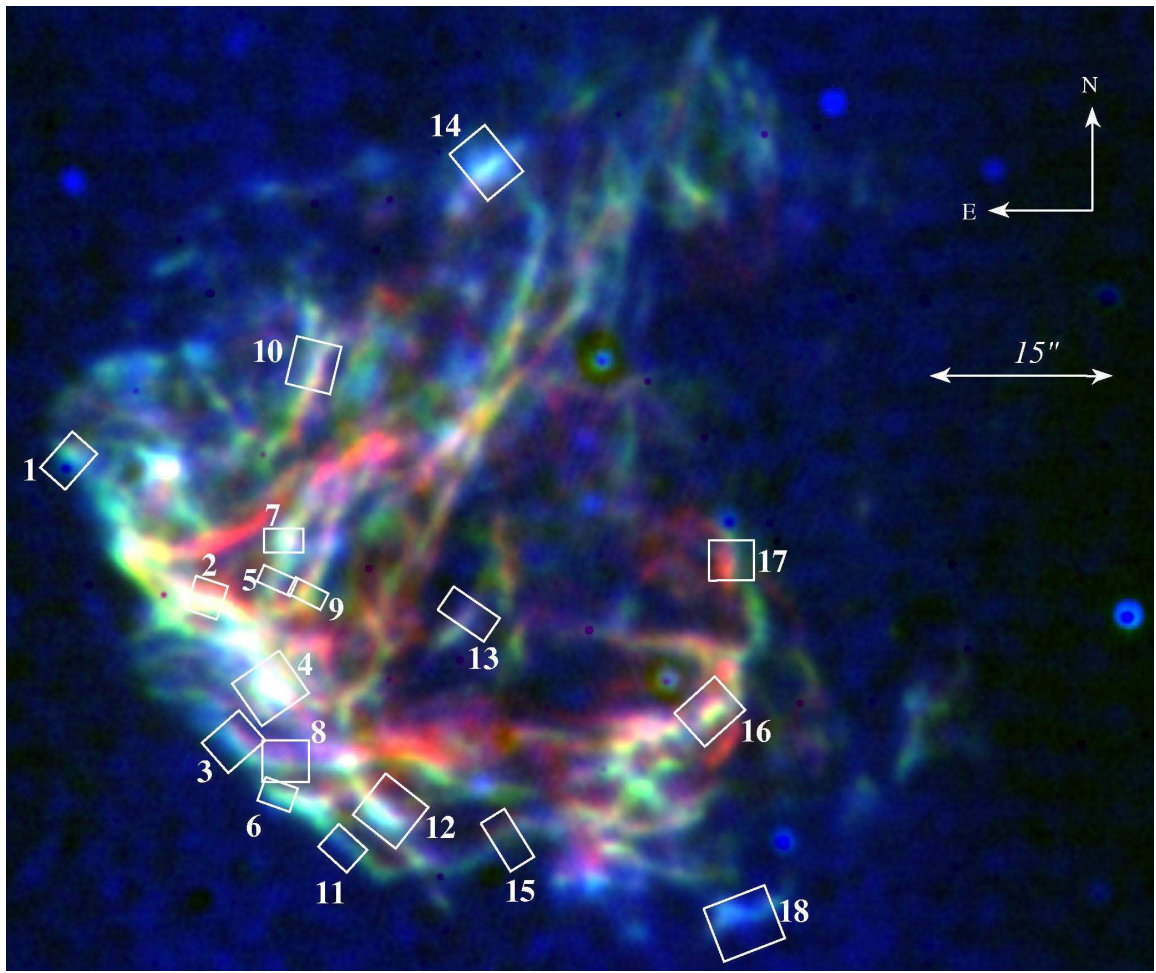}
\end{center}
\figcaption{Three-color image of N49 in \ha\ (\textsl{red}), 
\oiii\ (\textsl{green}), and \nev\ (\textsl{blue}). The \nev\ image
was first binned by 4 pixels then smoothed with a 1.38\arcsec\
Gaussian kernel. \ha\ and \oiii\ were not binned but were smoothed
to 0.345\arcsec . 
Regions from which the filament profiles were extracted are indicated.
\label{3color}}
\end{figure*}

Our data were calibrated with observations of the standard star
Hiltner~600 taken earlier in the night. The flux as a function of
wavelength was taken from LCALIB in IRAF, in particular the flux of
Hiltner~600 at 3426\AA\ is  
4.35731$\times 10^{-13}$ erg cm$^{-2}$ s$^{-1}$ \AA$^{-1}$ at the top
of the atmosphere.
The extinction curves available in IRAF for the Cerro Tololo
Inter-American Observatory (CTIO) were used, as they are a reasonable
approximation to the extinction at Las Campanas. The filter
transmission curves had been measured at Omega Optical Inc., and
these data are available at Las Campanas.\footnote{The actual numbers were
  extracted from hard-copies of the transmission curves using a
  stand-alone version of ``Dexter'' from ADS.}
Over the \nev\ and UV continuum bandpasses the extinction as a
function of zenith distance was well fitted
by a third order polynomial, and the filter transmission by a
Gaussian. For other filters simple linear or quadratic interpolation
was preferred owing to features in the extinction curve or flat tops in
the transmission as a function of wavelength. All other aspects of the
throughput and detector efficiency were calibrated as a single
conversion factor using Hiltner~600. The aggregate efficiency in \nev\
was sadly only 4.9\%.
We estimate that our photometric uncertainty is around 30\% based on
the mildly variable seeing and the small differences between the
calibrated scaling factor for the continuum subtraction and the one
which best subtracted the stars.  Continuum subtracted images of N49
in \ha, \oiii, and \nev\ are displayed separately in
Figure~\ref{fuseap} and combined into a three-color image in
Figure~\ref{3color}: \ha\ (\textsl{red}),  \oiii\ (\textsl{green}),
and \nev\ (\textsl{blue}).

The absolute coordinate system was found by comparison to the
U.S. Naval Observatory (USNO) A2.0 catalogue
\citep{1998yCat.1252....0M} using WCSTools 
\citep{mink2002}. At least 23 matched stars were identified for each
filter with a mean radial offset of 0.\arcsec367. However the error
in the relative positions between the filters is a fraction of a pixel
(0.2 -- 0.75 pixel) since
that is based on registering the same stars in different frames
using {\texttt geomap} in IRAF.
We note that the X-ray knot identified by
\citet{2003ApJ...586..210P} is $\sim$5\arcsec\ south of the poorly subtracted
star on the western boundary of Figure~\ref{3color}.
\citet{2003ApJ...586..210P} suggest that this knot is a fragment of
ejecta from the SN explosion like those seen in the Vela SNR since it
lies outside the apparent boundary of N49 and spectral fits indicate
overabundances of O, Si and Fe relative to LMC values. There is
nothing at the position of the knot in the \nev, \ha, or \oiii\ images
that might be identified as cooler emission associated with the  X-ray knot.

\section{Discussion}

\subsection{Comparison of \nev\ and \ovi\ }

The only information previously available on plasma
at a few hundred thousand degrees in N49 comes from
measurements of the \ovi\ doublet with the Hopkins Ultraviolet
Telescope (HUT)
\citep{1992ApJ...394..158V}
and FUSE \citep{2000ApJ...538L..61B,2004AJ....128.1615S} 
UV spectrometers. These spectra provide important line profile
information, but they are subject to uncertainties in
reddening and in the effects of resonant scattering
by \ovi\ ions, both within the
emitting plasma \citep[e.g.][]{1992ApJ...400..214L,2003ApJ...584..770R}
and along the line of sight.  There is also a
question as to how much of the \ovi\ emission arises from
radiative shock waves in the 200 $\rm km~s^{-1}$ range,
and how much comes from 300--400 $\rm km~s^{-1}$ non-radiative
shocks that produce soft X-rays.  Both this question and
the uncertainty in the effects of radiative scattering
within the emitting gas can be partly answered by examining
the way the intensity is distributed within the apertures
used for the \ovi\ observation.

\begin{deluxetable}{lccccc}
\tablewidth{0pt}
\tablecaption{Observed Intensities ($10^{-13}~\rm erg~cm^{-2}~s^{-1}$) 
\label{measured}}
\tablehead{Aperture & \nev\ & \ha\ & \oiii\ & \ovi\tablenotemark{a} &
  WFPC \ha\tablenotemark{a}} 
\startdata
 X00502  & 2.19  & 210.  & 64.2 & 9.9  & 288.  \\
 C05502  & 1.69  & 64.4  & 28.0 & 6.2  & 99.0  \\
 C05503  & 1.39  & 103.  & 28.8 & 5.1  & 126.  \\
 C05501  & 0.34  & 12.3  & 6.60 & 0.96  & 20.0  \\
 HUT  & 1.30  & 91.5  & 33.6 & 6.0  &           
\enddata
\tablenotetext{a}{\citet{2004AJ....128.1615S}}
\end{deluxetable}

The FUSE and HUT apertures are superposed on each of the emission line
images of N49 in Figure~\ref{fuseap}. 
Table \ref{measured} gives the
measured intensities in the \ovi\ doublet, along with the intensities
in \nev\ $\lambda$3425, \ha, and \oiii\ $\lambda$5007 measured by
integrating over the apertures after masking out a few
poorly subtracted stars. Table~\ref{measured} also gives the H$\alpha$
intensities reported by \citet{2004AJ....128.1615S} extracted from an
archival HST image.

In Table \ref{deredden},
we give the intensities corrected for a mean reddening
E(B-V) = 0.37 as recommended by \citet{1992ApJ...394..158V}.
\citet{1992ApJ...394..158V} point out that the reddening is
probably not uniform over N49, and that a scatter in the
reddening value will introduce a correction factor
corresponding to a reduction in E(B-V). This is because some fraction
of the area within the aperture sees significantly less
reddening and contributes a disproportionate amount of flux.
\citet{2003ApJ...589..242S}
assume E(B-V) = 0.37 and that the
scatter in E(B-V) is Gaussian, with a width of 0.14, which
reduces the dereddening factor for \ovi\ by a factor of 0.4.
Sankrit et al. also estimate that resonance line scattering
within the remnant
\citep{1992ApJ...400..214L,2003ApJ...584..770R}
attenuates the \ovi\ flux by about a factor
of 1.5, and that scattering in the halo of the LMC decreases
the flux by another 20\%.  The resonance scattering partly
compensates for the effect of scatter in E(B-V), and the
net result is to reduce the corrected flux of \ovi\ by 28\%.
We include this factor in the fluxes given in Table \ref{deredden}.  


\begin{deluxetable}{lccccc}
\tablewidth{0pt}
\tablecaption{Dereddened Intensities ($10^{-13}~\rm erg~cm^{-2}~s^{-1}$) 
\label{deredden}}
\tablehead{Aperture & \nev\ & \ha\ & \oiii\ & \ovi\ & WFPC \ha\ } 
\startdata
 X00502 & 13.6 & 549. & 221. & 2180. & 752.  \\
 C05502 & 10.5 & 168. & 96.4 & 1360. & 258.  \\
 C05503 & 8.66 & 269. & 99.2 & 1120. & 329.  \\
 C05501 & 2.13 & 32.1 & 22.7 & 212. & 52.2  \\
 HUT & 8.10 & 239. & 115. & 1320.  &       
\enddata
\end{deluxetable}

The first important comparison is the \nev/\ovi\ ratio.
We have computed a model with an updated version
of the steady flow shock radiative code of \citet{1979ApJS...39....1R}
and the LMC abundance set of  \citet{1992ApJ...401..220V}.
The ratio of \ovi\
to \nev\ depends only weakly on shock speed in this range, and
because the cooling is so heavily dominated by O and Ne at
temperatures of 3-6$\times 10^5$ K, only the relative abundances
of O and Ne come into play. The Vancura et al. LMC abundance
set gives O/Ne = 5.0.  The major limitation of these models
is their neglect of the thermal instabilities that affect shocks
faster than about 150 $\rm km~s^{-1}$ 
\citep[e.g.,][]{1992A&A...256..660I,2003ApJ...591..238S}.  
However, the \nev\ and \ovi\ lines are 
formed at such similar temperatures that thermal instabilities should
not change their ratio significantly. The predicted intensity ratio of
\nev\ to \ovi\ ranges from 0.0090 to 0.011 for shocks in the 200
to 300 $\rm km~s^{-1}$ range.  The ratios derived from Table \ref{deredden}
range from 0.0062 to 0.010, in very good agreement with the
predictions. This confirms that the reddening and resonance line
scattering correction factors of \citet{1992ApJ...394..158V} 
are reliable.

There is one major difference between the \ovi\ and \nev\ formation
temperatures.  The emissivity of \ovi\ has a high temperature tail 
extending to about $2 \times 10^6$ K, while \nev\ does not.  
Hence enhanced \ovi\ might indicate some emission from higher
temperature shocks. In fact, the smallest \nev/\ovi\ ratio is in
FUSE aperture X00502 where the X-rays are particularly bright. 
However, the uncertainties in the measurements, the reddening correction,
the relative abundances of Ne and O, and the model calculations can
easily account for the observed \nev/\ovi\ ratios which lie about 20\%
below those expected from the radiative shock models. 
Thus the
agreement of the observed flux ratio with the predictions
indicates that the bulk of the \ovi\ emission arises from
200--300 km s$^{-1}$ radiative shocks as opposed to
the non-radiative shocks that are responsible for producing the X-ray
emission from N49.
%

The second question is the degree of structure of the intermediate
temperature gas within the FUSE apertures.  
The \nev\ image in Figure \ref{fuseap} shows that in each of the
30\arcsec\ $\times$ 30\arcsec\ apertures most of the emission 
comes from filaments a few arcseconds across that cover perhaps 1/3
of the area within the aperture.  For example, in the X00502 aperture
the three brightest knots covering less than 2\% of the aperture
contribute 1/8 of the flux. 
This is what one would expect 
from the above result that the \ovi\ arises from radiative shocks.
 When it is possible to obtain an actual \ovi\ image,
it should show less contrast than does the \nev\ image, because
scattering will selectively dim the bright, edge-on filaments
in resonance lines \citep{1992ApJ...395L...9C}. 

The image of N49 gives a total \nev\ luminosity of the remnant 
of $1.2\times 10^{36}~\rm erg~s^{-1}$.  From the \nev/\ovi\ ratios,
this implies $L_{\mathrm O~VI} = 1.5 \times 10^{38}~\rm erg~s^{-1}$.  
This is 8 times the X-ray luminosity measured by \citet
{1981ApJ...248..925L}($1.9\times 10^{37}$ erg s$^{-1}$ from 0.15--4.5 keV).
According to the models described above, the \ovi\ luminosity
is about 10\% of the total energy dissipated by a shock in the 200--300
$\rm km~s^{-1}$ range.  Thus the luminosity of these intermediate
velocity shocks is around $10^{39}~\rm erg~s^{-1}$ and in the 
6,600 yr estimated lifetime of N49 \citep{2003ApJ...586..210P} 
they would have radiated $2 \times
10^{50}$ erg, a significant fraction of the nominal energy of a
supernova.  According to the shock speed distribution of
\citet{1992ApJ...401..220V},  slower shocks
also produce a comparable luminosity.  However, if the SNR has recently
reached the shell of a bubble produced by the progenitor
\citep{1985MNRAS.212..799S} or the dense
cloud to the East and South~\citep{1991supe.conf..679H} 
after passing rapidly through low
density gas, the age of the shocks could be significantly
less than 6,600 yr.  Indeed, \citet{2003ApJ...586..210P} find a
time-scale as short as 100 years based on the ionization state of the
X-ray emitting gas in the SE.
It is also worth noting that $L_{\mathrm O~VI}$ is 2.5 times the
infrared (IR) luminosity attributed to thermal emission from dust 
\citep{1987ApJ...319..126G}. \citet{2006astro.ph..7598W} suggest 
that much of this IR
emission comes from emission lines rather than dust, in which case it
is produced by intermediate velocity or slower shocks.

For comparison, estimates of the \ovi\ luminosity of the
Cygnus Loop range from $1.2 \times 10^{36}~\rm erg~s^{-1}$
based on a rocket observation \citep{1992ApJ...396L.103R}
to  $4.6 \times 10^{36}$ erg~s$^{-1}$ based on a Voyager
Ultraviolet Spectrometer observation
\citep{1991ApJ...374..202B}. 
These estimates would be reduced by a factor of 2 for the recent
distance estimate of 540 pc \citep{2005AJ....129.2268B}.
Thus the O VI luminosity of N49 is 50 to 100 times that of the Cygnus Loop.

\citet{1992ApJ...396L.103R} find
that the blast wave can account for the \ovi\ luminosity of the Cygnus
Loop, while \citet{1993ApJ...417..663V} %
find that the blast wave accounts for
only 10 to 15\% of the \ovi\ photons.  
 \citet{1992ApJ...396L.103R}
point out the overall morphological resemblance between X-ray and \ovi\
emission, while \citet{2001AJ....122..938D}
discuss a region in the Cygnus Loop where intermediate velocity shocks
clearly dominate.  The \nev\ data favor the interpretation
of intermediate velocity shocks for N49.  The blast wave temperatures
in N49 exceed 0.4 keV \citep{2003ApJ...586..210P}, which is too high
to produce O VI in the high temperature tail.  There could
in principle be substantial quantities of gas at 1 to $2 \times 10^6$ K
whose X-ray emission would be overwhelmed by that of the hotter
plasma.  Indeed, the detection of [\ion{Fe}{14}] 
\citep{1979ApJ...231L.147D}
requires some gas at these temperatures.  However, the
\nev\ to \ovi\ ratio derived above shows that this hot gas
does not dominate the \ovi\ luminosity.

\subsection{Spatial distribution and flux ratios of \nev, \oiii, and \ha}

\citet{1992ApJ...394..158V} 
noted the existence of a rim of \oiii\ extending outside of
the \ha\ emission. They interpreted this as a photoionized
precursor. Our images (as well as HST) show that the \oiii\ emission
actually \textit{peaks} outside of \ha.
However, based on the
fraction of \oiii\ emission in the narrow spike in the \oiii\ echelle
profiles \citet{1992ApJ...394..158V} find that the precursor
accounts for less than 20\% of the \oiii\ emission. 
Thus the exterior \oiii\ filaments must be from the radiative
shock itself.  This suggests a physical separation between the \oiii\ and
\ha\ in the post-shock gas. In Figure \ref{3color} we can identify 
many exterior and interior
filaments where there appears to be a distinct separation between
\oiii\ and \ha.

Efficiently radiating intermediate velocity shocks are widely thought
to be subject to thermal instabilities for $v_{s} > 150$ km s$^{-1}$,
but magnetic \citep{1992A&A...256..660I} or cosmic-ray
\citep{2006A&A...452..763W} pressure can suppress or damp the instability. 
Large variations in the line fluxes and ratios and wide separations
between the emitting regions of different ions may indicate thermally
unstable shock conditions. 
In the non-steady-state radiative shock models of
\citet{1992A&A...256..660I}, that include the ``cushioning'' effect of
magnetic pressure support on the collapsing thermally overstable filaments, a
peak in \oiii\ emission ahead of the Balmer line emission was
indicative of the presence of secondary shocks. Under the 
particular shock speed and ambient conditions they chose
($V_{S}=175$~km~s$^{-1}$, $n_{0}=1$~cm$^{-3}$, $B_{0}=3\mu$G) 
the separations between peaks are typically 4--7\arcsec $\times 
(D/1$~kpc$)^{-1}$
but can at times be as large as 20\arcsec  $\times (D/1$~kpc$)^{-1}$. 
\citet{1992A&A...256..660I} also finds that 
the crossings of multiple secondary shocks result in relatively
short-term time variations in both the line-intensities and
line-ratios with line-ratio fluctuations of greater than a factor of 20. 
They further predict an \oiii\ photoionization precursor extending
45\arcsec $\times (D/1$~kpc$)^{-1}$ with intensities 15--20\% of the peak
\oiii\ emission which can account for the echelle spectra.

At the distance to
the LMC the calculations of \citet{1992A&A...256..660I} translate into
separations as large as 0.4\arcsec. 
However the densities
required to explain the flux are at least a factor of 20 higher than
modeled by \citet{1992A&A...256..660I} while the lower LMC abundances will
increase the cooling time by a factor of 3. Thus we would
expect the separations to be a factor of 6 smaller than the
predictions of \citet{1992A&A...256..660I} unless there was
significantly more pressure support either from a higher magnetic
field or cosmic-rays. 

%
%
%

To quantify the separations and flux variations, 
18 regions were chosen along ``straight'' sections of
filaments where relatively clean profiles could be extracted in each
filter. These include the clearest examples of spatial separations plus 
a few profiles to represent the more typical cases where the peaks in all
bands essentially coincide.
The profiles, labeled ``1'' to ``18'' from east to west in Figure
\ref{3color}, were fit with Gaussians to 
determine the peak amplitude, position and full-width-half-maximum
(FWHM) of the filament in each band above a flat background level. 
Of these, 13 profiles were well-characterized by this simple model,
while others with wider profiles or that simply reached a plateau were not.
Selected profiles are shown in Figure \ref{profile} and plots of the
relative distances between peaks, the ratios of the integrated fluxes,
and the relative widths of the filaments are shown in Figure
\ref{plots}. From this investigation we find the following:

\begin{figure*}
\begin{center}
\noindent\includegraphics[width=2.75in]{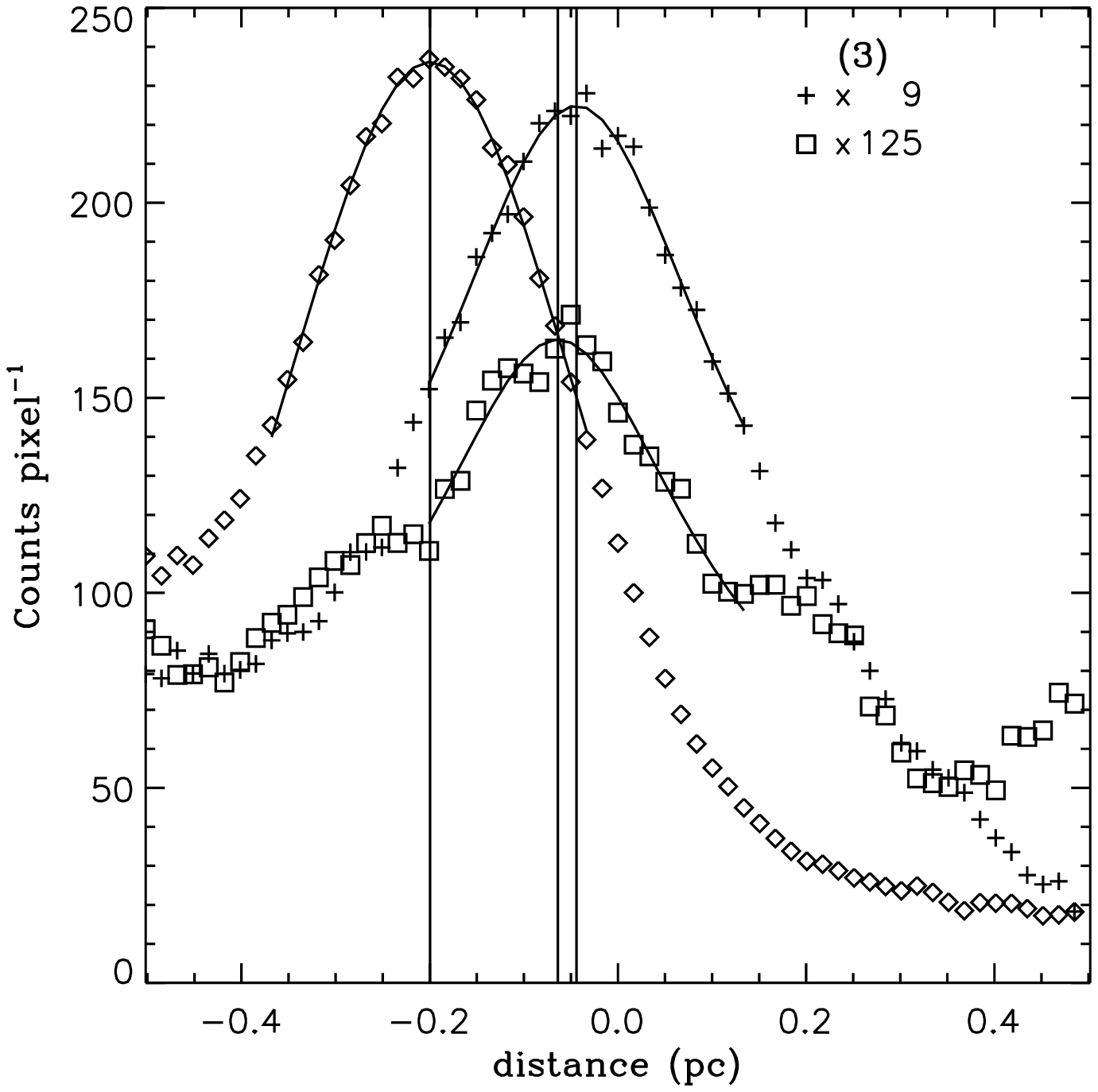} \hspace{0.2in} 
\noindent\includegraphics[width=2.75in]{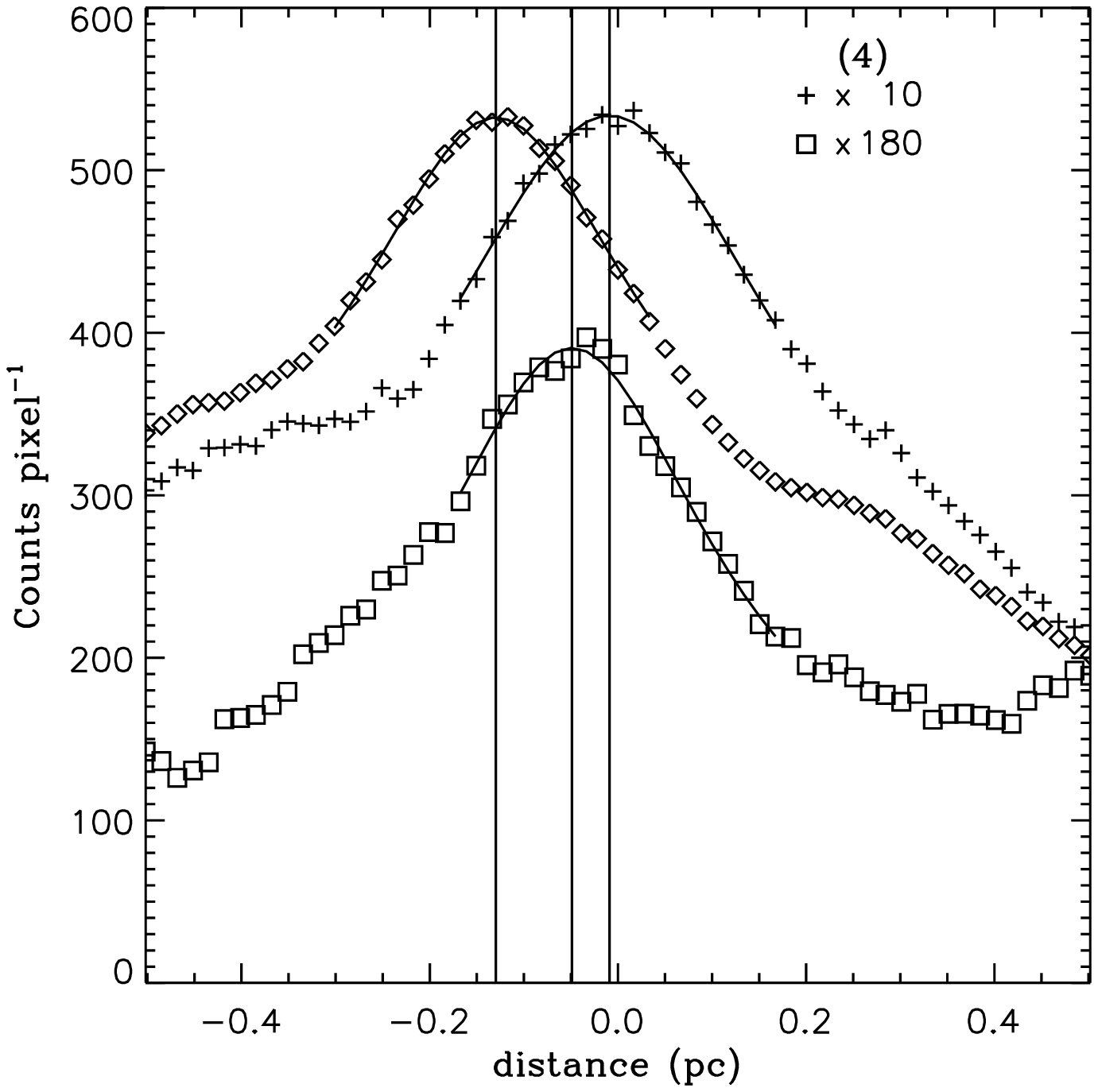} \\ 		  
\noindent\includegraphics[width=2.75in]{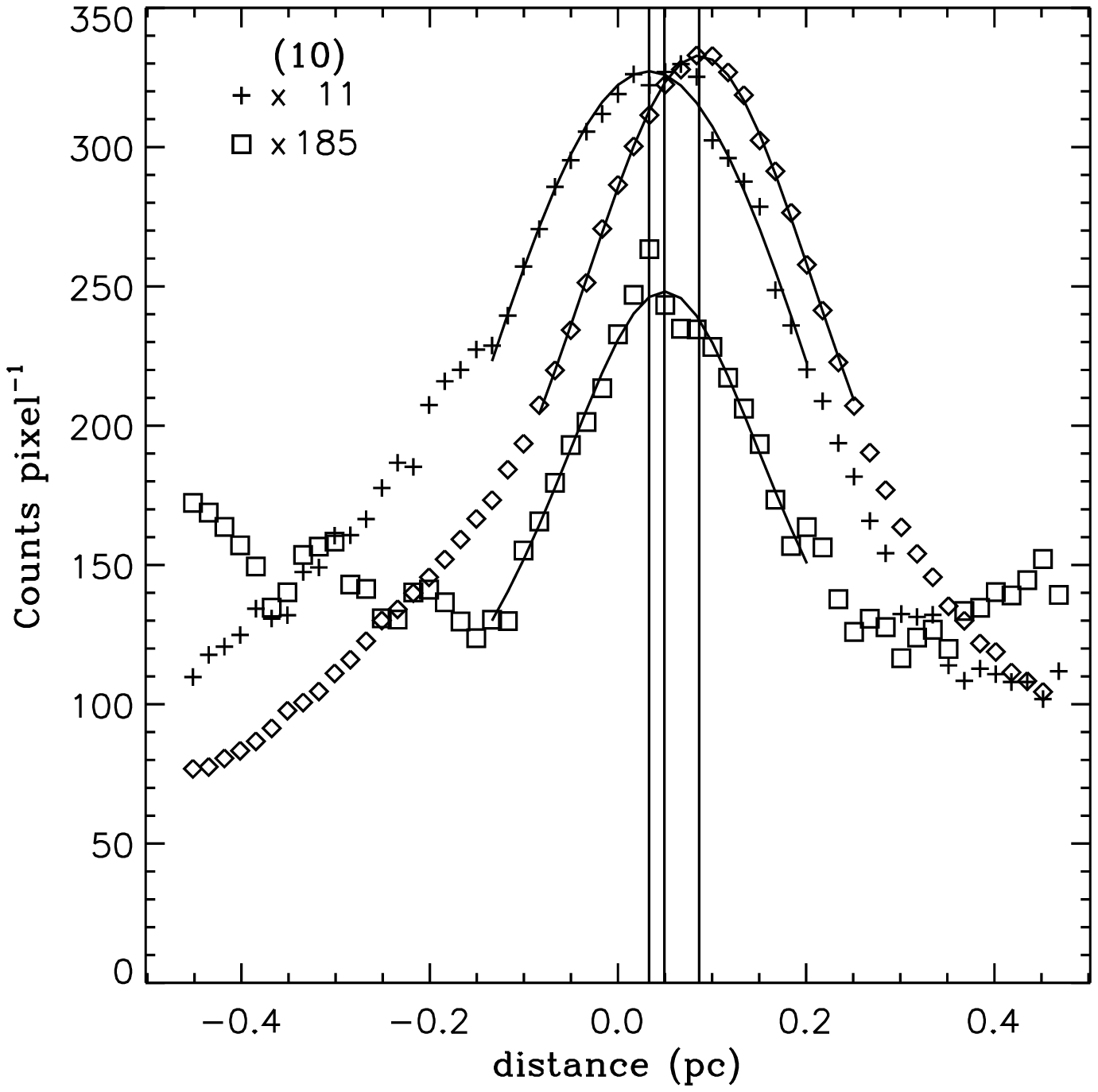}\hspace{0.2in} 
\noindent\includegraphics[width=2.75in]{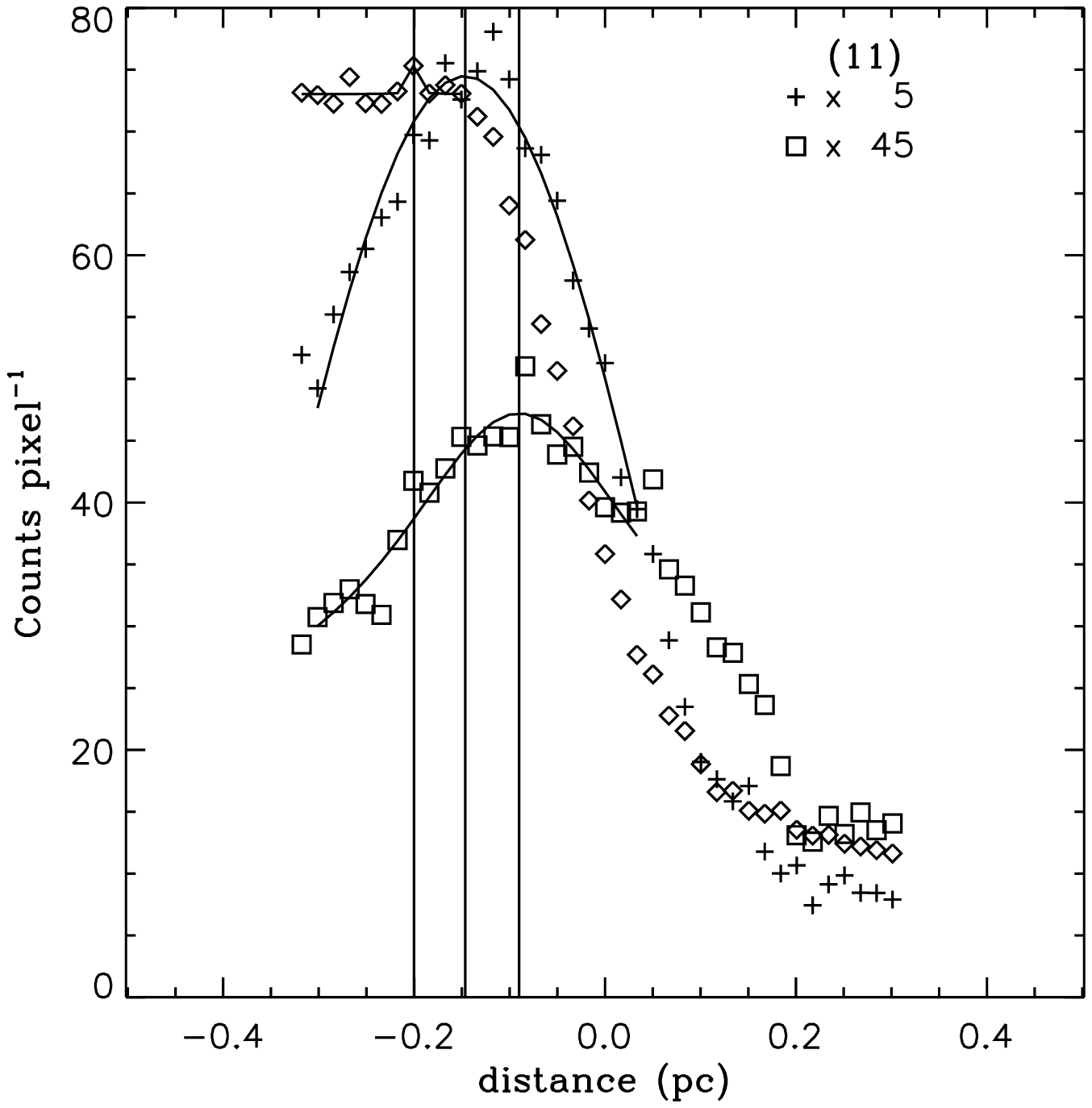} \\            
\noindent\includegraphics[width=2.75in]{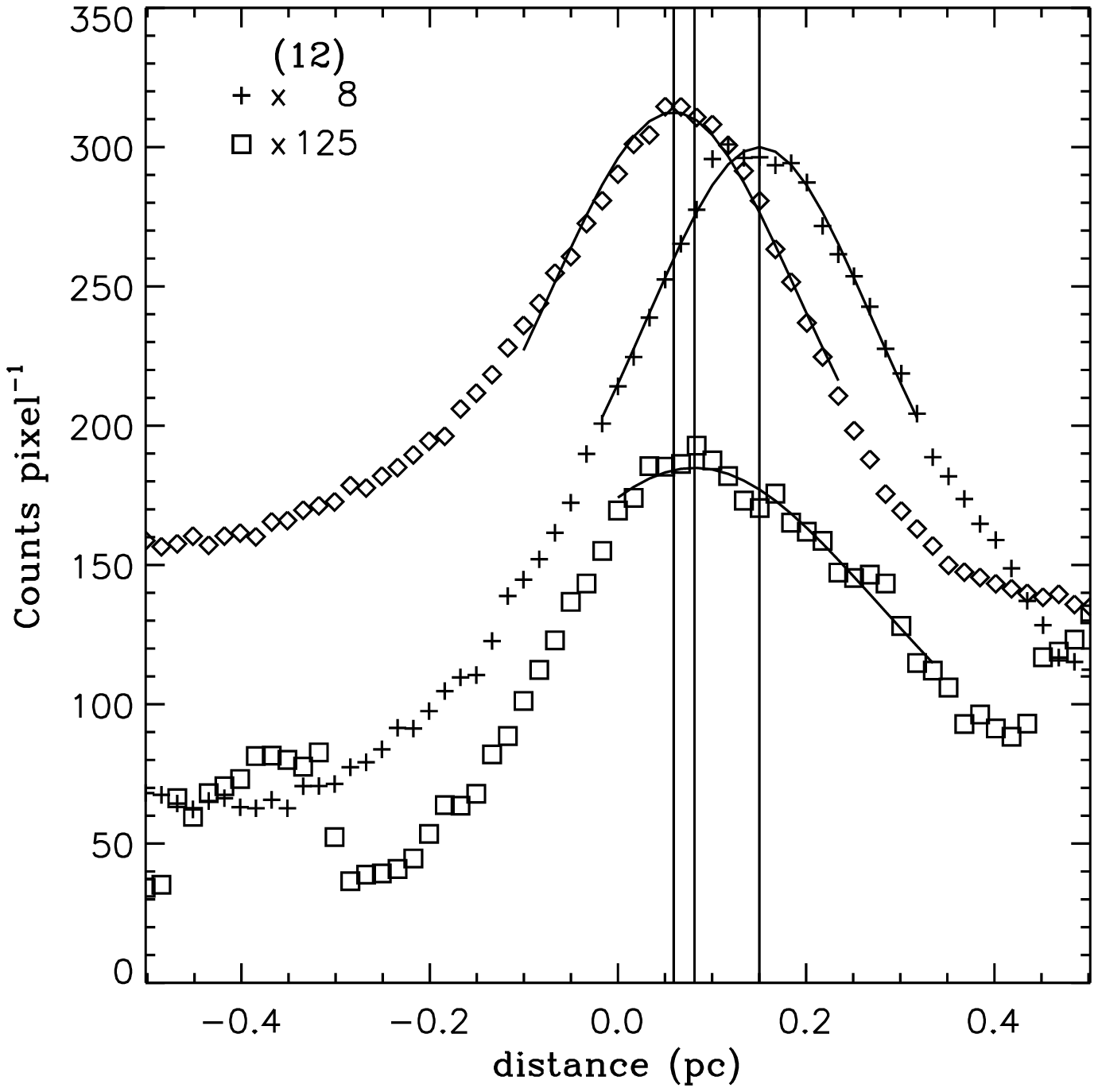} \hspace{0.2in}
\noindent\includegraphics[width=2.75in]{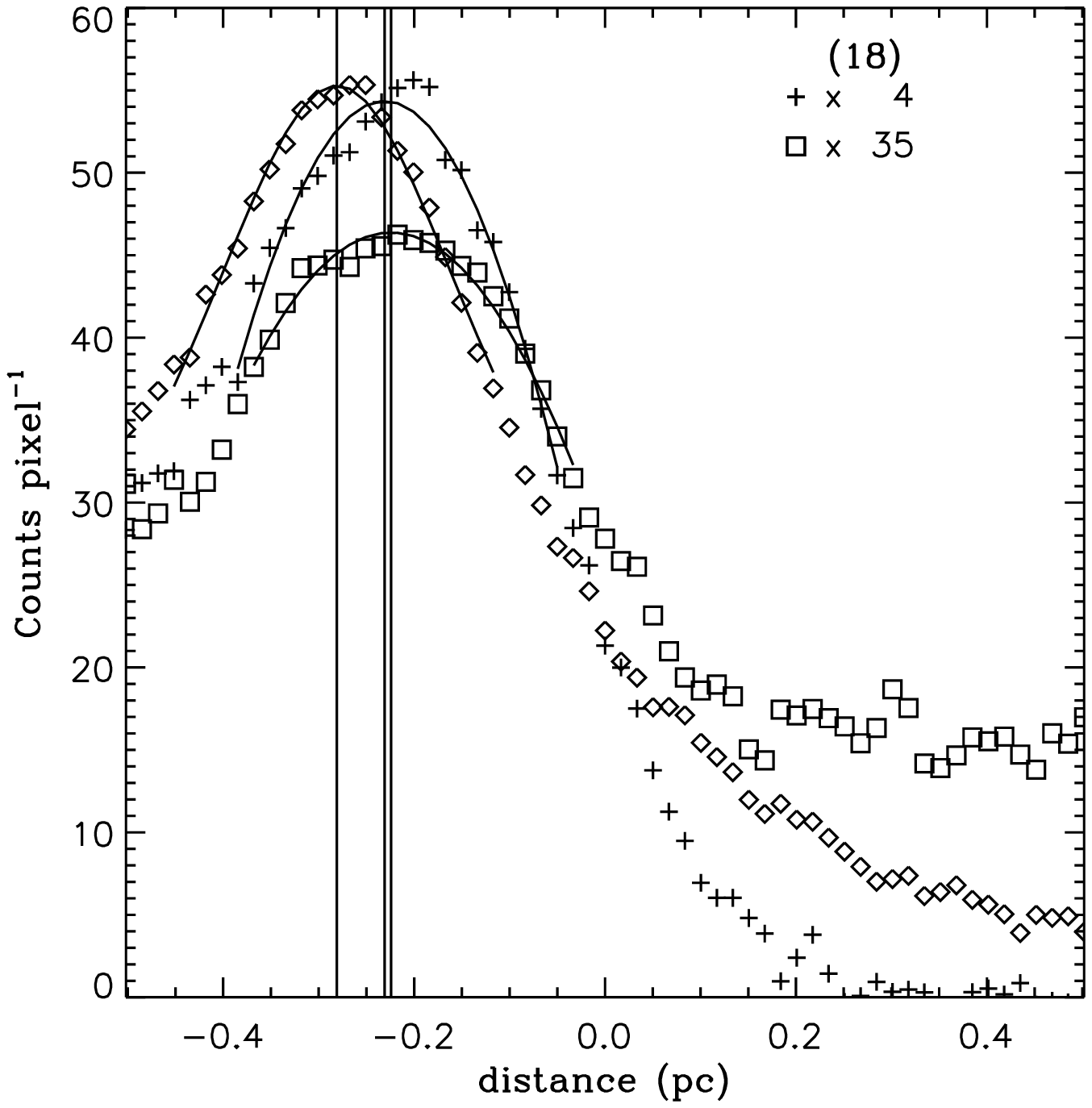} 
\end{center}
\figcaption{Intensity profiles across selected filaments, \ha\
  (\textsl{diamonds}), \oiii\ (\textsl{pluses}), and \nev\
  (\textsl{squares}). Six profiles 3,4,10,11,12, and 18, were chosen
  to demonstrate the range of properties seen among the 18
  extracted. The fainter \oiii\ and \nev\ have been multiplied by the
  factors shown in each plot. All profiles 
  were extracted from the original unsmoothed, unbinned data and are
  displayed in terms of the average counts pixel$^{-1}$ versus the
  distance in pc to give a better sense of the statistics. 
  Conversion factors from counts to flux and the exposure times 
  are as given in Table \ref{observe}.
\label{profile}}
\end{figure*}

\begin{itemize}

\item
{Any given ratio of fluxes \nev / \ha, \nev / \oiii, or \oiii / \ha\
  varies over  2 orders of magnitude (as measured by
  the parameters of the Gaussian fits), with the highest \nev / \ha\
ratio occuring in region 18 at the far southwestern edge of the
remnant. }

\item{In most of the regions the widths of the filament in all bands
  are similar, but in regions  1, 6, 7, 11, 12, and 18 the
  \nev\ emission appears more extended. Region 11 (along the southeast
  bright portion)  in particular shows a wider profile in \nev\ 
  extending ahead of the other two. }

\item{Measured separations between the peak flux in \oiii\ and \ha\ are as
large as 0.15~pc (0.62\arcsec) in the extreme cases of regions 3 and
17, and at least 11 regions show separations larger than 0.05~pc (0.2\arcsec).
In over half 
of the regions the peak in \nev\ lies between the other two, closer to
the peak in \oiii .  The exceptions are regions 2 and 12 where \nev\
and \ha\ essentially coincide, and 5, 6, 11 and 14 where \nev\ leads
\oiii\ by as much as 0.04~pc (0.17\arcsec).}

\end{itemize}

The flux ratio variations are in line with the idea of different
filaments being at different stages of thermal instability.
In fact, we find that steady flow shock models cannot explain 
\oiii/\ha $>$2/3 or  \nev/\oiii $>$ 1/4 for LMC abundances, both 
of which are seen.
%
We also expect that the \oiii\ and \ha\ emission from cooler, compressed
gas should be narrower than the \nev\ filaments which trace the 
2--6$\times 10^{5}$K shocked gas.
We do see
evidence for this in some places. 
The positions of peak brightness
in \oiii, \ha, and \nev\ are often significantly offset from one
another by distances of order 0.1~pc
similar to the predictions for a much lower density,
solar abundance plasma \citep{1992A&A...256..660I}.
Thus if the flux ratios and separations in N49 are to be explained by thermal
instability significant magnetic or cosmic-ray pressure support would
be required to slow the progress of the instability.

\begin{figure}
\noindent\includegraphics[width=2.75in]{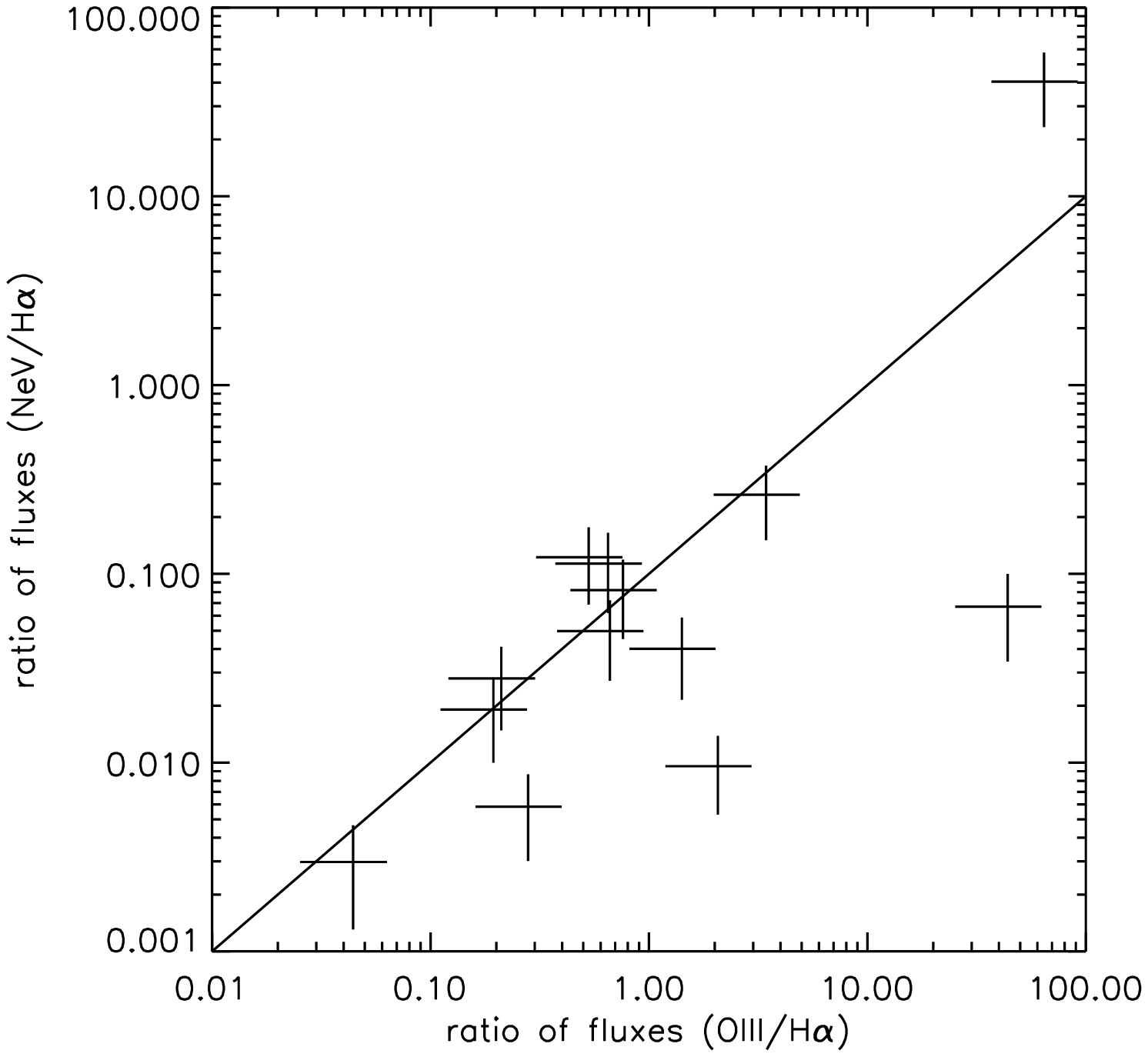} \\
\noindent\includegraphics[width=2.75in]{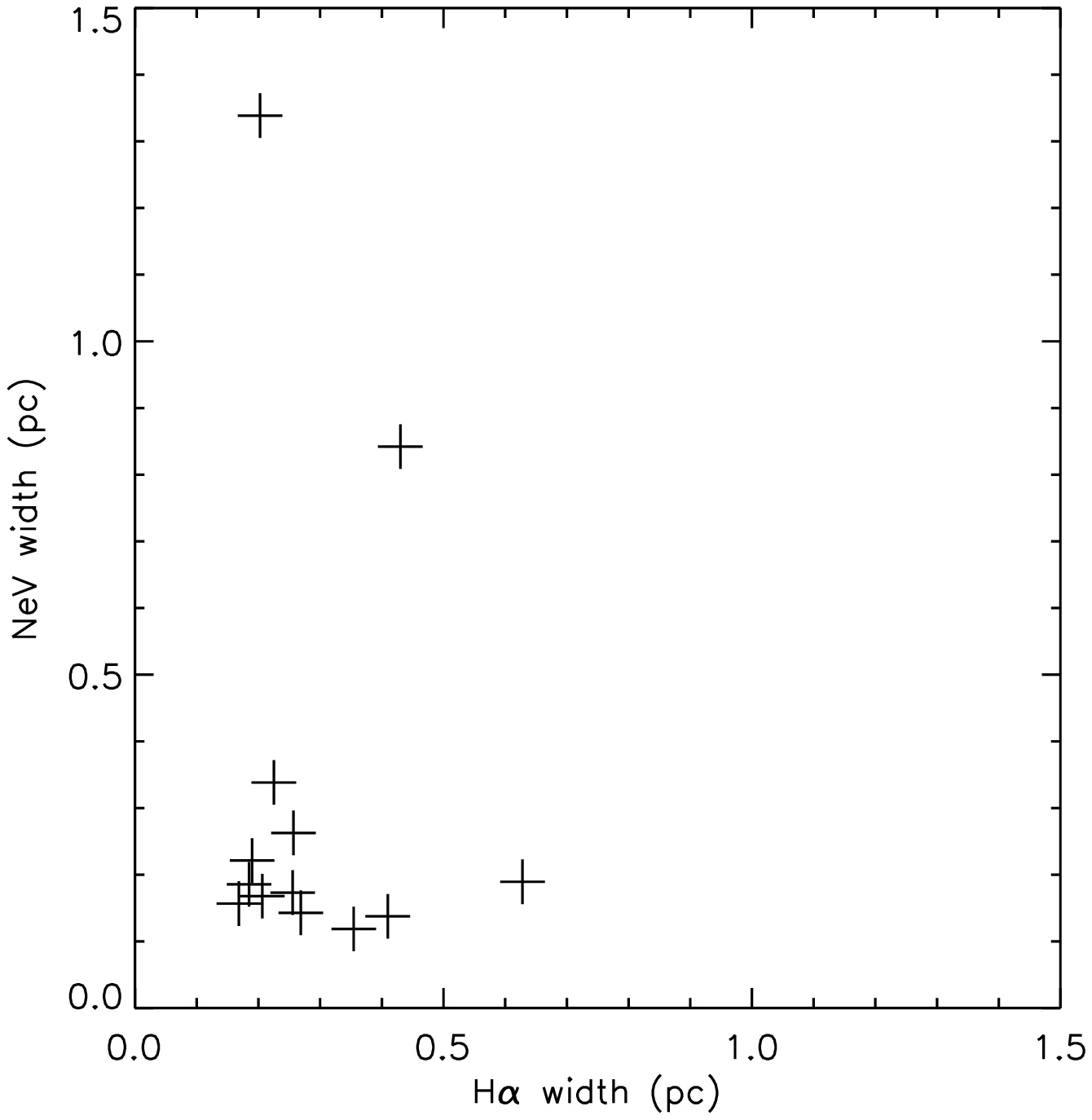} \\
\noindent\includegraphics[width=2.75in]{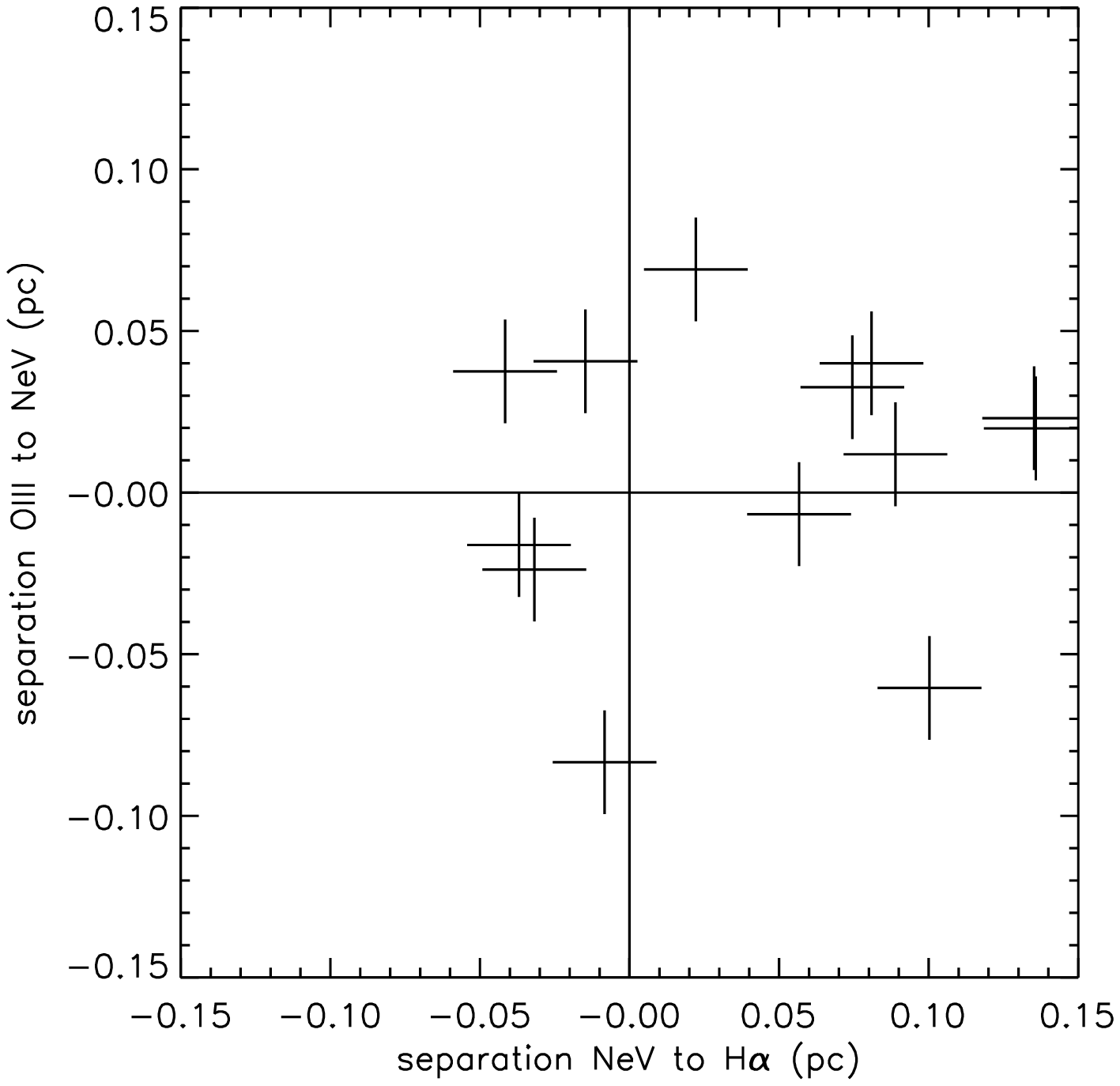}
\figcaption{Gaussian line-profile parameters for the selected
  regions. (a) Flux ratios of \nev/\ha\ vs. \oiii/\ha\ measured
  by the area under the Gaussian and dereddened. 
  (b) Width in pc of \nev\ vs. \ha .
  (c) Separation in pc, between the peaks of emission, \oiii\ to \nev\
  vs. \nev\ to \ha . Overplotted in (a) is the line \nev/\oiii\ = 0.1.
  Only regions for which a Gaussian line-profile
  captured the peak position and width were included in these plots. 
  Errors shown are based on (a) 30\% error in calibration and the
  square-root of the number of counts, (b) $\pm$2 pixels, and (c)
  $\pm$1 pixel. 
\label{plots}}
\end{figure}

The filamentary nature itself is evocative of 
thermally unstable shocks such as the ``ribbons of high density'' in
the simulations of \citet{1998ApJ...500..342B}. The
structures in N49 appear similar in scale and curvature to the late
times of their high ambient density simulation \citep[Figure
  11]{1998ApJ...500..342B}. Quantitatively an actual filamentary
structure of condensed ribbons is not required
because the surface brightness contrast between the filaments and the
diffuse emission (factors of 5--10 in  \oiii) is still consistent with a
rippled sheet geometry \citep{1987ApJ...314..187H}.
However the similarities are highly suggestive. 

\section{Summary}


N49 exhibits an extremely bright complex structure in \nev . The
radiative losses alone imply a recent interaction with 
dense material relative to the age of the remnant. 
The ratios of \nev\ to \ovi\ emission, which both occur at 
2--6$\times 10^{5}$K, are consistent with emission from
200--300~km~s$^{-1}$ shocks, confirming the dereddening relation of 
\citet{2003ApJ...589..242S} and implying that most of the \ovi\
emission arises from 200--300~km~s$^{-1}$ radiative shocks. 
For \nev, \oiii, and \ha, which occur at decreasing temperatures 
post-shock, we find $\sim$0.1~pc separations between the peaks of 
their emission in individual filaments and a factor of 100 
variation in their relative fluxes, including a number of places 
where the \oiii/\ha\ or \nev/\oiii\ flux ratios are higher than allowed 
by steady flow shock models at LMC abundances.
These properties are consistent with the presence of thermal
instabilities, but at the densities implied by the total flux,
additional magnetic or cosmic-ray pressure would be required.

Thermal instabilities have been almost universally predicted to occur in 
radiative shocks above $v_{s} > 150$~km~s$^{-1}$, 
however observational evidence for them has been 
severely lacking. Density fluctuations in the pre-shock medium can 
mask or mimic the effects of thermal instability and it can be 
difficult to distinguish between different departures from 
steady-flow. For instance 
in the Cygnus Loop, high \oiii/\ha\ flux ratios are interpreted 
as evidence for a truncated or incomplete recombination zone behind a recent 
shock interaction \citep{1980ApJ...238..881R,1988ApJ...324..869R}. 
In that particular region it seems unlikely that thermal instabilities 
have taken hold given the smooth variation of properties over such an extended 
thin structure \citep{1988ApJ...324..869R}.
In the slightly faster shocks of N49, on the other hand, we are 
interpreting the variability over the entire remnant as snapshots of 
different stages in thermally unstable shocks. 
In N49, the spatial separation and large variability are
the keys to interpreting the non-steady-flow flux ratios. However, without 
additional magnetic or cosmic ray pressure these structures would have 
been unresolvable.

\acknowledgments



The authors would like to thank Scott Burles for the use of his IDL
routine to identify the ``herringbone'' electrical noise using a
one-dimensional fast-fourier-transform and Matthew Holman for the 
valuable observation that in MagIC the pattern of electrical noise is
mirrored over each of the four chips allowing the diffuse source
emission to be removed prior to identifying the pattern.
C.E.R. was supported during this work by NASA Grant NAG5-9281.

{\it Facilities:} \facility{Magellan:Clay (MagIC)}

\bibliographystyle{apj}   
\bibliography{ms.bbl}

\begin{thebibliography}{34}
\expandafter\ifx\csname natexlab\endcsname\relax\def\natexlab#1{#1}\fi

\bibitem[{{Blair} {et~al.}(1991){Blair}, {Long}, {Vancura}, \&
  {Holberg}}]{1991ApJ...374..202B}
{Blair}, W.~P., {Long}, K.~S., {Vancura}, O., \& {Holberg}, J.~B. 1991, \apj,
  374, 202

\bibitem[{{Blair} {et~al.}(2005){Blair}, {Sankrit}, \&
  {Raymond}}]{2005AJ....129.2268B}
{Blair}, W.~P., {Sankrit}, R., \& {Raymond}, J.~C. 2005, \aj, 129, 2268

\bibitem[{{Blair} {et~al.}(2000){Blair}, {Sankrit}, {Shelton}, {Sembach},
  {Moos}, {Raymond}, {York}, {Feldman}, {Chayer}, {Murphy}, {Sahnow}, \&
  {Wilkinson}}]{2000ApJ...538L..61B}
{Blair}, W.~P., {Sankrit}, R., {Shelton}, R., {Sembach}, K.~R., {Moos}, H.~W.,
  {Raymond}, J.~C., {York}, D.~G., {Feldman}, P.~D., {Chayer}, P., {Murphy},
  E.~M., {Sahnow}, D.~J., \& {Wilkinson}, E. 2000, \apjl, 538, L61

\bibitem[{{Blair} {et~al.}(2002){Blair}, {Sankrit}, \&
  {Tulin}}]{2002ApJS..140..367B}
{Blair}, W.~P., {Sankrit}, R., \& {Tulin}, S. 2002, \apjs, 140, 367

\bibitem[{{Blondin} {et~al.}(1998){Blondin}, {Wright}, {Borkowski}, \&
  {Reynolds}}]{1998ApJ...500..342B}
{Blondin}, J.~M., {Wright}, E.~B., {Borkowski}, K.~J., \& {Reynolds}, S.~P.
  1998, \apj, 500, 342

\bibitem[{{Cornett} {et~al.}(1992){Cornett}, {Jenkins}, {Bohlin}, {Cheng},
  {Gull}, {O'Connell}, {Parker}, {Roberts}, {Smith}, {Smith}, \&
  {Stecher}}]{1992ApJ...395L...9C}
{Cornett}, R.~H., {Jenkins}, E.~B., {Bohlin}, R.~C., {Cheng}, K.-P., {Gull},
  T.~R., {O'Connell}, R.~W., {Parker}, R.~A.~R., {Roberts}, M.~S., {Smith},
  A.~M., {Smith}, E.~P., \& {Stecher}, T.~P. 1992, \apjl, 395, L9

\bibitem[{{Danforth} {et~al.}(2001){Danforth}, {Blair}, \&
  {Raymond}}]{2001AJ....122..938D}
{Danforth}, C.~W., {Blair}, W.~P., \& {Raymond}, J.~C. 2001, \aj, 122, 938

\bibitem[{{Dopita} \& {Mathewson}(1979)}]{1979ApJ...231L.147D}
{Dopita}, M.~A. \& {Mathewson}, D.~S. 1979, \apjl, 231, L147

\bibitem[{{Graham} {et~al.}(1987){Graham}, {Evans}, {Albinson}, {Bode}, \&
  {Meikle}}]{1987ApJ...319..126G}
{Graham}, J.~R., {Evans}, A., {Albinson}, J.~S., {Bode}, M.~F., \& {Meikle},
  W.~P.~S. 1987, \apj, 319, 126

\bibitem[{{Hester}(1987)}]{1987ApJ...314..187H}
{Hester}, J.~J. 1987, \apj, 314, 187

\bibitem[{{Hughes} {et~al.}(1991){Hughes}, {Bronfman}, \&
  {Nyman}}]{1991supe.conf..679H}
{Hughes}, J.~P., {Bronfman}, L., \& {Nyman}, L. 1991, in Supernovae, ed. S.~E.
  {Woosley} (Springer-Verlag, New York), 679

\bibitem[{{Innes}(1992)}]{1992A&A...256..660I}
{Innes}, D.~E. 1992, \aap, 256, 660

\bibitem[{{Korreck} {et~al.}(2004){Korreck}, {Raymond}, {Zurbuchen}, \&
  {Ghavamian}}]{2004ApJ...615..280K}
{Korreck}, K.~E., {Raymond}, J.~C., {Zurbuchen}, T.~H., \& {Ghavamian}, P.
  2004, \apj, 615, 280

\bibitem[{{Long} {et~al.}(1992){Long}, {Blair}, {Vancura}, {Bowers},
  {Davidsen}, \& {Raymond}}]{1992ApJ...400..214L}
{Long}, K.~S., {Blair}, W.~P., {Vancura}, O., {Bowers}, C.~W., {Davidsen},
  A.~F., \& {Raymond}, J.~C. 1992, \apj, 400, 214

\bibitem[{{Long} {et~al.}(1981){Long}, {Helfand}, \&
  {Grabelsky}}]{1981ApJ...248..925L}
{Long}, K.~S., {Helfand}, D.~J., \& {Grabelsky}, D.~A. 1981, \apj, 248, 925

\bibitem[{{Mink}(2002)}]{mink2002}
{Mink}, D.~J. 2002, in ASP Conf. Ser. 281: Astronomical Data Analysis Software
  and Systems XI, ed. D.~A. {Bohlender}, D.~{Durand}, \& T.~H. {Handley}, 169

\bibitem[{{Monet} {et~al.}(1998){Monet}, {Canzian}, {Dahn}, {Guetter},
  {Harris}, {Henden}, {Levine}, {Luginbuhl}, {Monet}, {Rhodes}, {Riepe},
  {Sell}, {Stone}, {Vrba}, \& {Walker}}]{1998yCat.1252....0M}
{Monet}, D.~B.~A., {Canzian}, B., {Dahn}, C., {Guetter}, H., {Harris}, H.,
  {Henden}, A., {Levine}, S., {Luginbuhl}, C., {Monet}, A.~K.~B., {Rhodes}, A.,
  {Riepe}, B., {Sell}, S., {Stone}, R., {Vrba}, F., \& {Walker}, R. 1998, 1252

\bibitem[{{Nishikida} {et~al.}(2006){Nishikida}, {Edelstein}, {Korpela},
  {Sankrit}, {Feuerstein}, {Min}, {Shinn}, {Lee}, {Yuk}, {Jin}, \&
  {Seon}}]{2006ApJ...644L.171N}
{Nishikida}, K., {Edelstein}, J., {Korpela}, E.~J., {Sankrit}, R.,
  {Feuerstein}, W.~M., {Min}, K.~W., {Shinn}, J.-H., {Lee}, D.-H., {Yuk},
  I.-S., {Jin}, H., \& {Seon}, K.-I. 2006, \apjl, 644, L171

\bibitem[{{Park} {et~al.}(2003){Park}, {Burrows}, {Garmire}, {Nousek},
  {Hughes}, \& {Williams}}]{2003ApJ...586..210P}
{Park}, S., {Burrows}, D.~N., {Garmire}, G.~P., {Nousek}, J.~A., {Hughes},
  J.~P., \& {Williams}, R.~M. 2003, \apj, 586, 210

\bibitem[{{Rasmussen} \& {Martin}(1992)}]{1992ApJ...396L.103R}
{Rasmussen}, A. \& {Martin}, C. 1992, \apjl, 396, L103

\bibitem[{{Raymond}(1979)}]{1979ApJS...39....1R}
{Raymond}, J.~C. 1979, \apjs, 39, 1

\bibitem[Raymond et al.(1980)]{1980ApJ...238..881R} Raymond, J.~C., 
Hartmann, L., Black, J.~H., Dupree, A.~K., \& Wolff, R.~S.\ 1980, \apj, 
238, 881 

\bibitem[Raymond et al.(1988)]{1988ApJ...324..869R} Raymond, J.~C., Hester, 
J.~J., Cox, D., Blair, W.~P., Fesen, R.~A., \& Gull, T.~R.\ 1988, \apj, 
324, 869 

\bibitem[{{Raymond} {et~al.}(2003){Raymond}, {Ghavamian}, {Sankrit}, {Blair},
  \& {Curiel}}]{2003ApJ...584..770R}
{Raymond}, J.~C., {Ghavamian}, P., {Sankrit}, R., {Blair}, W.~P., \& {Curiel},
  S. 2003, \apj, 584, 770

\bibitem[{{Sankrit} \& {Blair}(2002)}]{2002ApJ...565..297S}
{Sankrit}, R. \& {Blair}, W.~P. 2002, \apj, 565, 297

\bibitem[{{Sankrit} {et~al.}(2003){Sankrit}, {Blair}, \&
  {Raymond}}]{2003ApJ...589..242S}
{Sankrit}, R., {Blair}, W.~P., \& {Raymond}, J.~C. 2003, \apj, 589, 242

\bibitem[{{Sankrit} {et~al.}(2004){Sankrit}, {Blair}, \&
  {Raymond}}]{2004AJ....128.1615S}
---. 2004, \aj, 128, 1615

\bibitem[{{Sankrit} {et~al.}(2001){Sankrit}, {Shelton}, {Blair}, {Sembach}, \&
  {Jenkins}}]{2001ApJ...549..416S}
{Sankrit}, R., {Shelton}, R.~L., {Blair}, W.~P., {Sembach}, K.~R., \&
  {Jenkins}, E.~B. 2001, \apj, 549, 416

\bibitem[{{Sauvageot} {et~al.}(1999){Sauvageot}, {Decourchelle}, \&
  {Bohigas}}]{1999A&A...351..669S}
{Sauvageot}, J.~L., {Decourchelle}, A., \& {Bohigas}, J. 1999, \aap, 351, 669

\bibitem[{{Shull} {et~al.}(1985){Shull}, {Dyson}, {Kahn}, \&
  {West}}]{1985MNRAS.212..799S}
{Shull}, Jr., P., {Dyson}, J.~E., {Kahn}, F.~D., \& {West}, K.~A. 1985, \mnras,
  212, 799

\bibitem[{{Sutherland} {et~al.}(2003){Sutherland}, {Bicknell}, \&
  {Dopita}}]{2003ApJ...591..238S}
{Sutherland}, R.~S., {Bicknell}, G.~V., \& {Dopita}, M.~A. 2003, \apj, 591, 238

\bibitem[{{Szentgyorgyi} {et~al.}(2000){Szentgyorgyi}, {Raymond}, {Hester}, \&
  {Curiel}}]{2000ApJ...529..279S}
{Szentgyorgyi}, A.~H., {Raymond}, J.~C., {Hester}, J.~J., \& {Curiel}, S. 2000,
  \apj, 529, 279

\bibitem[{{Vancura} {et~al.}(1992{\natexlab{a}}){Vancura}, {Blair}, {Long},
  {Davidsen}, {Bowers}, {Dixon}, {Durrance}, {Feldman}, {Ferguson}, {Henry},
  {Kimble}, {Kriss}, {Kruk}, \& {Moos}}]{1992ApJ...401..220V}
{Vancura}, O., {Blair}, W.~P., {Long}, K.~S., {Davidsen}, A.~F., {Bowers},
  C.~W., {Dixon}, W.~V.~D., {Durrance}, S.~T., {Feldman}, P.~D., {Ferguson},
  H.~C., {Henry}, R.~C., {Kimble}, R.~A., {Kriss}, G.~A., {Kruk}, J.~W., \&
  {Moos}, H.~W. 1992{\natexlab{a}}, \apj, 401, 220

\bibitem[{{Vancura} {et~al.}(1992{\natexlab{b}}){Vancura}, {Blair}, {Long}, \&
  {Raymond}}]{1992ApJ...394..158V}
{Vancura}, O., {Blair}, W.~P., {Long}, K.~S., \& {Raymond}, J.~C.
  1992{\natexlab{b}}, \apj, 394, 158

\bibitem[Vancura et al.(1993)]{1993ApJ...417..663V} Vancura, O., Blair, 
W.~P., Long, K.~S., Raymond, J.~C., \& Holberg, J.~B.\ 1993, \apj, 417, 663 

\bibitem[{{Wagner} {et~al.}(2006){Wagner}, {Falle}, {Hartquist}, \&
  {Pittard}}]{2006A&A...452..763W}
{Wagner}, A.~Y., {Falle}, S.~A.~E.~G., {Hartquist}, T.~W., \& {Pittard}, J.~M.
  2006, \aap, 452, 763

\bibitem[{{Williams} {et~al.}(2006){Williams}, {Chu}, \&
  {Gruendl}}]{2006astro.ph..7598W}
{Williams}, R.~M., {Chu}, Y.~., \& {Gruendl}, R. 2006, \aj, in press

\end{thebibliography}



\end{document}